\documentclass[iop]{emulateapj_2012_alexander}

\bibpunct{}{}{,}{a}{}{;}

\begin{document}

\author{J. Alexander and S. Gulyaev}
\title{On The Apparent Narrowing of Radio Recombination Lines at High Principal Quantum Numbers}

\affil{Institute for Radio Astronomy and Space Research, Auckland University of Technology, Auckland, New Zealand}

\shorttitle{Apparent Narrowing of RRLs at High $n$}
\shortauthors{Alexander \& Gulyaev}


\received{5 August 2011}
\accepted{14 November 2011}

\begin{abstract}
We critically analyze the Bell et al. findings on ``anomalous''
widths of high-order Hydrogen radio recombination lines in the Orion
Nebula at 6 GHz. We review their method of modified frequency switching
and show that the way this method is used for large $\Delta n$ is
not optimal and can lead to misinterpretation of measured spectral
line parameters. Using a model of the Orion Nebula, conventional broadening
theory, and Monte Carlo simulation, we determine a transition zone
$n=224,\,\ldots,\,241$ ($\Delta n=11,\,\ldots,\,14$), where measurement errors
grow quickly with $n$ and become comparable with the measurement
values themselves. When system noise and spectrum channelization are
accounted for, our simulation predicts ``processed'' line narrowing
in the transition zone similar to that reported by Bell et al. We
find good agreement between our simulation results and their findings,
both in line temperatures and widths. We conclude, therefore, that
Bell et al.'s findings do not indicate a need to revise Stark broadening
theory.
\end{abstract}

\section{Introduction}

In the 1990s, Morley Bell and coauthors developed a technique for
measuring weak spectral lines by reducing broad baseline variations
(\citet{Bell1997}). This technique was referred to as ``modified
frequency switching'' and was used to detect of weak atomic and molecular/maser
lines in spectra of gaseous nebulae, circumstellar envelopes, and
star formation regions (see, e.g., \citet{Bell&Feldman1991,Belletal1992,Belletal1993}).

Bell and coauthors used this technique to measure radio recombination
line (RRL) widths and temperatures (\citet{Belletal2000}, hereafter
BASV). At 6 GHz, they found that the ``...{[}processed H{]} \emph{lines
at large n are both narrower and stronger than expected from theory}...''
and suggested, ``\emph{This behavior is ... inconsistent with Griem's
}theory ...''. This publication, with its subtitle: ``Confrontation
with theory at high principal quantum numbers'' induced a wave of
publications where BASV's finding was called an ``anomaly'', ``puzzle''
and even ``mystery''. For example, \citet{Oks2004} paper is titled
``On the puzzle of the observed narrowing of radio recombination
lines,'' \citet{Griem2005} concludes his paper by writing that the
result ``...remains a mystery...,'' \citet{Gavrilenko&Oks2007}
calls the result a ``dramatic discrepancy,'' and \citet{Watson2006}
concludes his abstract with ``Thus this mystery is not resolved by
the present calculations''.

While some authors have sought an explanation of these findings in
the revision of Stark broadening theory (e.g., \citet{Oks2004,Watson2006}),
some remain skeptical about the modified frequency switching technique
or suggest that it requires verification. For example, \citet{vonProchetal2010}
writes ``...it is easy to distort the RRL line shape using data reduction
techniques (Bell et al. 2000).'' \citet{Griem2005} referred to private
communication with Bell stressing that ``...measurement errors related
to large reduction factors from unprocessed to processed Voigt profile
widths can probably not be excluded (M. B. Bell 2004, private communication)''.
In their book, \citet{Gordon&Sorochenko2009} concluded: ``The Bell
et al. (2000) results are so different from what had been expected,
and the observing technique is so new, that we suggest waiting for
an independent confirmation of the observations before accepting a
fault in the present theory of RRL Stark broadening.''

BASV's findings are based on frequency switched or ``processed''
observational data, which were recently presented in \citet{Belletal2011}
and re-interpreted in \citet{Bell2011}. Their findings result from
recursive frequency switching, in software, of observational data
that were initially recorded at the telescope using hardware frequency switching. 

In this paper, we investigate the frequency switching technique and
demonstrate that, if applied correctly, it has a number of advantages.
Frequency switching removes many gain variations and does not require
subjective estimates of the zero level of a spectrum. Though the technique
helped BASV to detect RRLs with $\Delta n$ greater than $\Delta n=6$
(\citet{Smirnovetal1984}), the way the method was used was not optimal
and, as such, cannot be used to test the theory of spectral line broadening.
We present simulations based on the \citet{Lockman&Brown1975} model
of the Orion Nebula and conventional theory of spectral line broadening.
We apply observational specifications from BASV and \citet{Belletal2011},
including frequency range, channel width, frequency switching offset,
number of frequency switching overlaps, and noise temperature rms.
Results of our simulation demonstrate good agreement with BASV's findings,
both in line width and temperature. The computed ``processed'' widths
exhibit narrowing similar to that reported in BASV. We show that BASV's
spectral line ``narrowing'' is the result of the way the observational
data were processed and that BASV's findings do not contradict the
existing Stark broadening theory.

\section{Software Frequency Switching}\label{sec:Software-Frequency-Switching}

In this section, we introduce the data reduction technique used by
BASV to systematically acquire information about weak spectral features
in the presence of baseline fluctuations. \citet{Bell1997} established
and named this technique SOFMOR (``small-offset frequency-switching
multiple overlap reduction''). This technique was then called ``modified
frequency switching'' (MFS) in BASV. SOFMOR and MFS are frequency
switching techniques that use a frequency offset fixed in the receiver
hardware (see, e.g., \citet{Robinson1964}), followed by recursive frequency
switching in software. Here, we refer to this technique (whether conducted
in hardware or software) as FS (frequency switching). 

In FS, the original spectrum is subtracted from a copy which is shifted
by a number of channels (an offset). Mathematically, it can be related
to a finite difference iteration.

If the offset is $h$, then the notation 
\begin{equation}
f_{p}=f(x+p\, h)
\end{equation}
 can be used. Formulae for the results of $m$ FS iterations ($m$
finite differences) are then 
\begin{eqnarray*}
f^{(1)} & = & f_{1}-f_{0}\\
f^{(2)} & = & f_{2}-2f_{1}+f_{0}\\
f^{(3)} & = & f_{3}-3f_{2}+3f_{1}-f_{0}\\
f^{(4)} & = & f_{4}-4f_{3}+6f_{2}-4f_{1}+f_{0}
\end{eqnarray*}
 and so on (see, e.g., \citet{Beyer1987}, p. 449; \citet{Zwillinger2002},
p. 705), therefore 
\begin{equation}
f^{(m)}=\sum_{k=0}^{m}f_{k}{m \choose k}(-1)^{m+k}\quad.\label{finite_diffs}
\end{equation}

When $f(x)$ is given with uncertainty $\pm\sigma_{0}$, where $\sigma_{0}$
is standard deviation, the resulting uncertainty in $f^{(m)}$ can
be derived from the error propagation rule (see, e.g., \citet{Taylor1997},
p. 75) as 
\begin{equation}
\sigma_{m}=\sigma_{0}\sqrt{\sum_{k=0}^{m}\left(\frac{\partial f^{(m)}}{\partial f_{k}}\right)^{2}}=\sigma_{0}\sqrt{\sum_{k=0}^{m}{m \choose k}^{2}}=\sigma_{0}\sqrt{{2m \choose m}}.\label{sigma}
\end{equation}
 Given that an FS spectrum overlaps with itself, the case of \emph{dependent}
samples must be applied (see below).

If $f(x)$ is a single channel feature (digital analogue of $\delta$-function),
that is 
\[
f(x)=\delta(x_{0})=\left\{ \begin{array}{ll}
1, & \;\; x=x_{0}\\
0, & \;\;{\rm {otherwise,}}
\end{array}\right.
\]
 following Equation (\ref{finite_diffs}), $m$ FS-overlaps generates a series
of equidistant peaks with amplitudes ${m \choose k}(-1)^{m+k}$. This
is illustrated in the following Pascal's triangle for six consecutive
overlaps or three cycles: 

\begin{table}[!h]
{\small
\addtolength{\tabcolsep}{-3pt}
\begin{tabular}{ccccccccccccccc}

overlap, \textit{m}&&&&&&&&&&&&&&cycle, \textit{i}\\
0 &  &  &  &  &  &  & 1 &  &  &  &  &  &  & 0\\
1 &  &  &  &  &  & -1 &  & 1\\
2 &  &  &  &  & 1 &  & -2 &  & 1 &  &  &  &  & 1\\
3 &  &  &  & -1 &  & 3 &  & -3 &  & 1\\
4 &  &  & 1 &  & -4 &  & 6 &  & -4 &  & 1 &  &  & 2\\
5 &  & -1 &  & 5 &  & -10 &  & 10 &  & -5 &  & 1\\
6 & 1 &  & -6 &  & 15 &  & -20 &  & 15 &  & -6 &  & 1 & 3

\end{tabular}}.
\end{table}

Each cycle of two overlaps results in a symmetric pattern with a prominent
central feature. The amplitude of the central feature after $i$ FS-cycles
is the central binomial coefficient: 
\begin{equation}
|A_{i}|={2i \choose i}=2,\;6,\;20,\;{\rm \dots}\hspace{0.7cm}(i=1,\;2,\;3,{\rm \dots})\quad.
\end{equation}

According to Equation (\ref{sigma}), the standard deviation after $i$ FS-cycles
\begin{equation}
\sigma_{i}=\sqrt{{4i \choose 2i}}=\sqrt{6},\;\sqrt{70},\;\sqrt{924},\;{\rm \dots}\hspace{0.7cm}(i=1,\;2,\;3,\;4,{\rm \dots})\;,\label{sigma2n}
\end{equation}
 resulting in a signal-to-noise ratio (S/N) of 
\begin{equation}
{\rm S/N}_{i}=\frac{{2i \choose i}}{\sqrt{{4i \choose 2i}}}=0.82,\;\;0.72,\;\;0.66,\;{\rm \dots}\hspace{0.7cm}(i=1,\;2,\;3,{\rm \dots})\;.\label{SNRSNR}
\end{equation}
 Using Stirling's formula (see, e.g., \citet{Abramovich1965}) we obtain
\begin{equation}
{\rm S/N}_{i}\approx\left(\frac{2}{\pi i}\right)^{1/4}=0.89\; i^{-1/4}\quad.\label{SNR1}
\end{equation}
 Therefore, BASV's S/N tends to \emph{decrease}
with the number of FS-cycles, $i$, as $i{}^{-1/4}$ .

When \emph{independent} samples are superimposed, the standard deviation
is scaled as $\sigma_{i}=2,\;4,\;8,\ldots$ , which results in the
growth of the S/N: ${\rm S/N}_{i}={2i \choose i}/2^{i}=1,\;1.5,\;2.5,\;\ldots\;$
($i=1,\;2,\;3,\;\ldots$). This common approach is not applicable
to the FS technique used in BASV. Instead, they overlap the once recorded
spectrum with itself and therefore the case of dependent samples must
be used, as derived in Equations (\ref{sigma2n})--(\ref{SNR1}).

Though the first overlap in BASV occurs in hardware, where two independent
samples are mixed, the S/N remains unchanged because
the first overlap of two samples produces a $\sqrt{2}$ increase in
the resulting standard deviation regardless of whether the two samples
are dependent (software FS) or independent (hardware FS). 

\begin{figure}[h]
\includegraphics[bb=30bp 75bp 567bp 580bp,clip,scale=0.47]{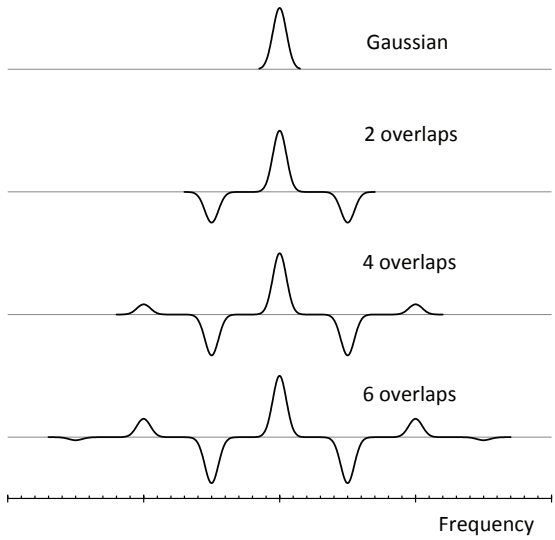}
\caption{Transformation of a Gaussian shape after 2, 4, and 6 FS-overlaps. The
initial width, $w$, is much less than the FS offset, $h$; $w/h=0.2$.
This plot is a graphical depiction of Pascal's triangle in Section
\ref{sec:Software-Frequency-Switching}. To compare the shapes and
widths, we invert (where necessary) the resulting spectra and normalize
the peak intensity of the central feature to the intensity of the
initial Gaussian. }

\label{fig:gauss_2_4_6overlaps} 
\end{figure}

\begin{figure}[h]
\includegraphics[bb=30bp 75bp 567bp 560bp,clip,scale=0.47]{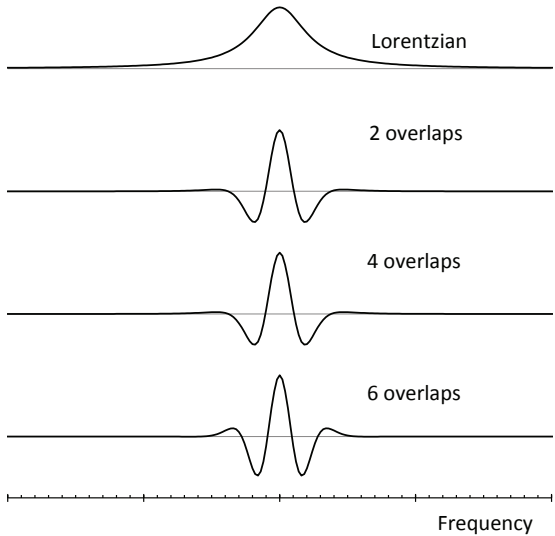}
\caption{Transformation of a Lorentzian shape after 2, 4, and 6 FS overlaps.
The initial width, $w$, is greater than the offset, $h$. The ratio
$w/h=4$ is used, which corresponds to BASV's $\Delta n=20$ case.
To compare the shapes and widths, we invert (where necessary) the
resulting spectrum and normalize the peak intensity of the central
feature to the intensity of the initial Lorentzian.}

\label{fig:lorentz_2_4_6overlaps} 
\end{figure}

Figure \ref{fig:gauss_2_4_6overlaps} shows a Gaussian feature characterized
by its full width at half-maximum (FWHM), $w$, and its transformation
after 2, 4, and 6 FS-overlaps are applied to the initial feature. The
frequency offset or FS-offset, \emph{h}, is chosen to be much greater
than the width, $h\gg w$. Additional features appear offset from
the central feature, whose offsets are multiples of $h$. All the
features are shifted relative to the position of the original feature,
$x_{0}$. After \emph{i} FS-cycles, the central feature is located
at $x=x_{0}-i\, h$. These shifts are removed in Figures \ref{fig:gauss_2_4_6overlaps}
and \ref{fig:lorentz_2_4_6overlaps}, such that all features are centered
vertically. Additionally, all features are normalized and inverted
(where necessary) to compare their shapes and widths. 

If $h\gg w$, the additional features (referred to as ``reference
images'' in \citet{Bell1997}) can be removed by adding or subtracting
scaled versions of the central feature to them---a process referred
to as ``cleaning'' in \citet{Bell1997}. For example, after two
FS-overlaps, the additional features shown in Figure \ref{fig:gauss_2_4_6overlaps}
can be removed by adding the central feature divided by 2. Dividing
this result by $-2$ recovers the original feature exactly. After
four overlaps, shown in Figure \ref{fig:gauss_2_4_6overlaps}, the
additional features are removed by adding $-1/6$ and $4/6$ of the
central feature to them. Dividing the central feature by 6 recovers
the original feature exactly. 

This normalization procedure keeps the central feature's amplitude
unchanged regardless of the number of FS-cycles, $i$. However, according
to Equation (\ref{SNR1}), the noise level grows as $i{}^{1/4}$ . 

\begin{figure}[h]
\includegraphics[bb=30bp 120bp 567bp 780bp,clip,scale=0.47]{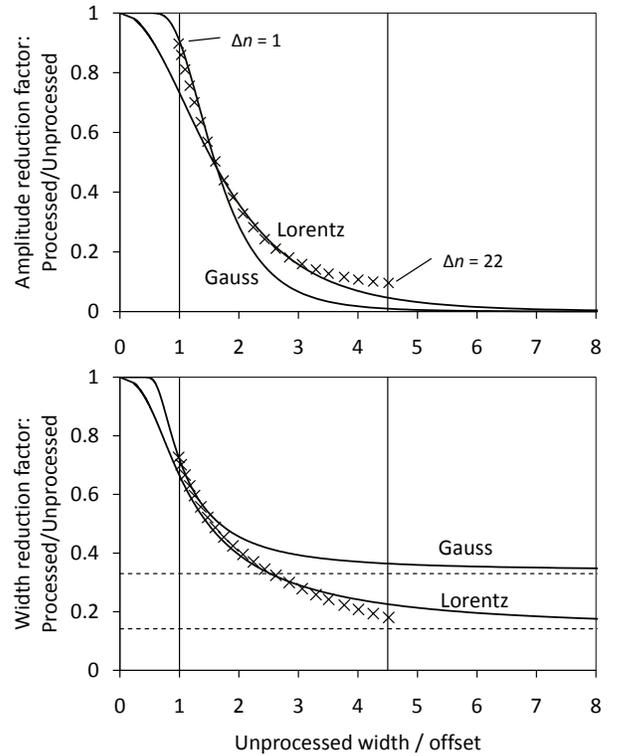}
\caption{Reduction factors (processed over unprocessed) for the peak line temperature
(top plot) and the line width (bottom plot) as a function of the normalized
unprocessed width, $w/h$. Both plots are calculated for six FS-overlaps.
Vertical lines show minimal and maximal values of $w/h$ for BASV's
observations: $1\lesssim w/h\lesssim4.5$. Crosses show reduction
factors predicted by the three-component Orion Nebula model of \citet{Lockman&Brown1975}. Horizontal dashed lines in the bottom plot are
the asymptotes for Gaussian and Lorentzian curves as $w/h\rightarrow\infty$.
These plots demonstrate that, for a fixed FS-offset \emph{h}, significant
reduction in spectral line intensity and width results from FS as
the unprocessed line width increases due to Stark broadening. }

\label{fig:amps_widths_6_overlaps} 
\end{figure}

NRAO's 140 ft telescope was used by BASV to observe the Orion Nebula
at 6 GHz. Based on the 140 ft data archive of the 1992 April observation
(integration time of $\sim\!48$ hours, system temperature of $\sim\!120\,\textrm{K}$,
and channel width of $\sim\!78\,\textrm{kHz}$), a noise temperature
$T_{\textrm{rms}}\approx1\,\textrm{mK}$ could be achieved. Given
that BASV used six overlaps (one hardware and five software FS-overlaps)
or three cycles, Equation (\ref{SNR1}) implies an ``frequency-switched'' temperature
rms of $T_{\textrm{rms}}\approx1.5\,\textrm{mK}$.

Figure \ref{fig:lorentz_2_4_6overlaps} illustrates the case when
the FS-offset is less than the original line width, $h<w$, specifically
$h=w/4$. This figure shows the transformation of a Lorentzian feature
after 2, 4, and 6 FS-overlaps. To compare the shapes and widths, we
normalize the peak intensity of the central feature to the intensity
of the initial feature. Figure \ref{fig:lorentz_2_4_6overlaps} shows
that the central feature of the processed line is much narrower than
the width of the original Lorentzian. Given that FS processing reduces
line widths and amplitudes when $h<w$, BASV refer to them as ``processed''
widths and amplitudes. 

The Voigt profile, which models the RRL shape, is a convolution of
Gaussian and Lorentzian profiles, where the Gaussian represents the
Doppler-broadened contribution and the Lorentzian represents the impact-broadened
contribution (\citet{Gordon&Sorochenko2009}). Figure \ref{fig:amps_widths_6_overlaps}
shows how the amplitudes and widths of FS-processed Gaussian and Lorentzian
profiles vary as a function of the normalized unprocessed width, $w/h$.
These plots show amplitude and width reduction factors (processed
to unprocessed ratio) calculated for six FS-overlaps. The vertical lines
show the limits of BASV's experiment: $1\lesssim w/h\lesssim4.5$.
Crosses are reduction factors predicted by the \citet{Lockman&Brown1975}
Orion Nebula model (see the next section). Horizontal dashed lines in
the bottom plot are the asymptotes for Gaussian and Lorentzian cases
as $w/h\rightarrow\infty$. 

\begin{figure}[h]
\includegraphics[bb=130bp 60bp 797bp 530bp,clip,scale=0.385]{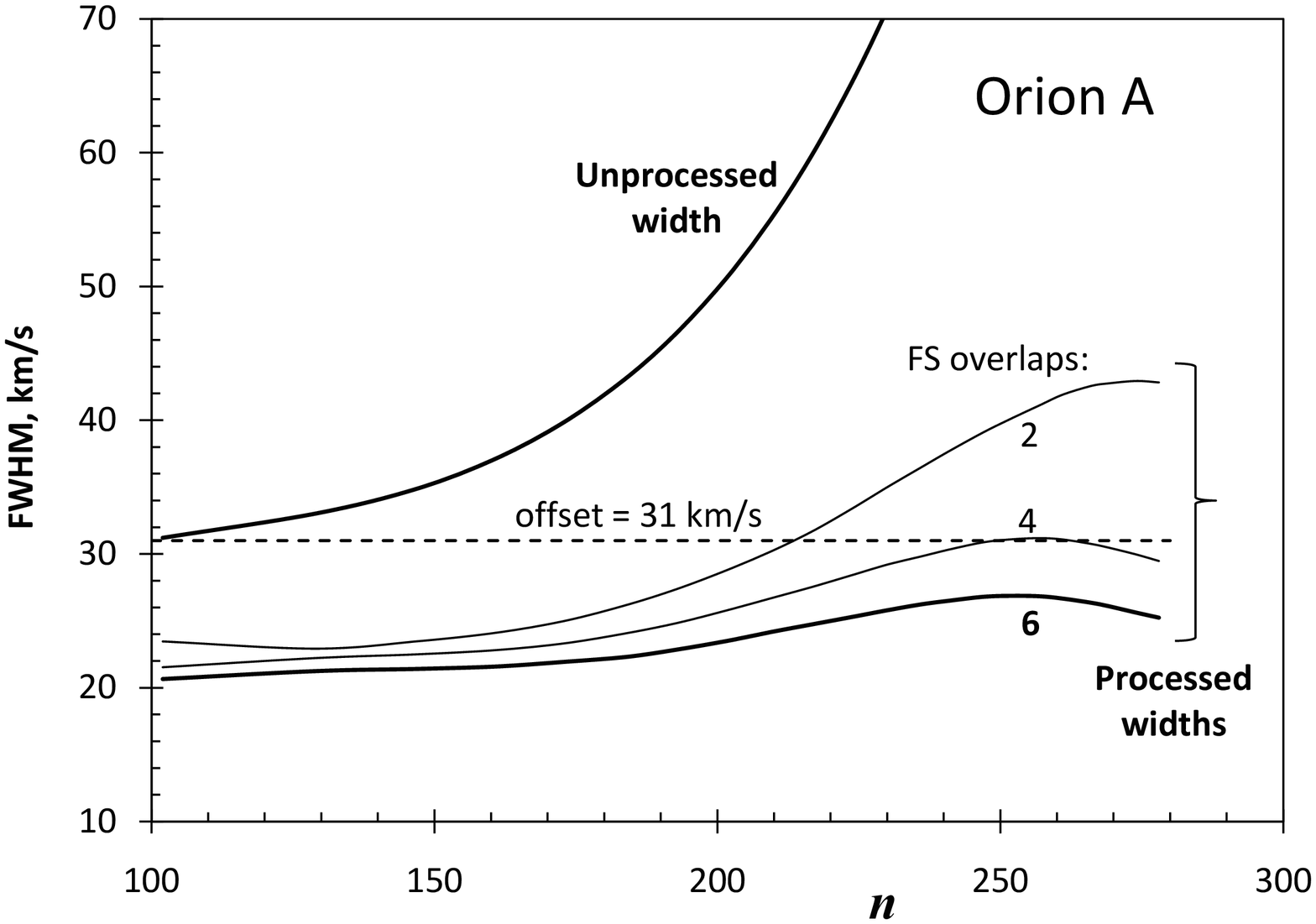}
\caption{Processed width vs. $n$ for different number of FS overlaps at 6
GHz. Simulation is conducted for the Orion Nebula model of \citet{Lockman&Brown1975};
no noise is taken into account. Dashed line shows BASV's FS-offset
(eight channels = 31.2 km s$^{-1}$). The upper (thick)
curve shows the unprocessed width $w$ vs. $n$. The lower curve corresponds
to the BASV's case of six FS-overlaps. This plot demonstrates that when
$h<w$, the processed width is \emph{increasingly} insensitive to
the significantly changing unprocessed width as the number of overlaps
increase. This indicates that the FS technique in BASV's case is a
poor probe of line broadening.}

\label{fig:width_processed_versus_n_and_overlaps_orion_6ghz} 
\end{figure}

\section{Application of FS to Model Spectra}

The frequency switching technique used in BASV consists of two independent
steps: (1) overlapping the spectrum and (2) ``cleaning'' the overlapped
spectrum. Overlapping six times creates a set of \textquotedblleft{}reference
images\textquotedblright{} for each line feature in the spectrum.
The result of overlapping (step 1) is illustrated in Figures \ref{fig:gauss_2_4_6overlaps}
$(h\gg w$) and \ref{fig:lorentz_2_4_6overlaps} ($h<w$). When $h<w$
(BASV's case), overlapping reduces the line widths to below the FS-offset,
$h$. 

After overlapping, the reference images are removed. In Bell (1997)
and BASV, this ``cleaning'' procedure was done manually, one line
at a time, starting with the strongest line and working to the weakest.
BASV's reported line widths after cleaning are in the range of 18
to 30 km s$^{-1}$, which is less than or about the Doppler width and less
than the FS-offset (Figures 2 and 4 in BASV). The conclusion made
in BASV about line \textquotedblleft{}narrowing\textquotedblright{}
is based on interpretation of these \textquotedblleft{}processed\textquotedblright{}
widths. 

A multi-component non-LTE radiative transfer simulator was created
to explore the effect of FS on model RRL spectra. For the Orion Nebula
we use the three-component model of \citet{Lockman&Brown1975}, which
simulates the beam size of the 140 ft radio telescope. The model consists
of a compact (0.043 pc) dense ($N_{e}=10^{4.5}\,\textrm{cm}{}^{-3}$)
symmetric core located behind two extended (0.56 and 2.50 pc) layers
of gas of lower density ($N_{e}=10^{3.5}$ and $10^{2.3}\,\textrm{cm}{}^{-3}$).
The corresponding temperatures are $T_{e}=$7500 K (core), 10,000 K,
and 12,500 K (outer layer). This model predicts continuum emission
and H$n\alpha$ and H$n\beta$ spectra in agreement with observations
between 1 and 90 GHz (\citet{Lockman&Brown1975}).

\begin{figure}[h]
\includegraphics[bb=130bp 60bp 797bp 530bp,clip,scale=0.380]{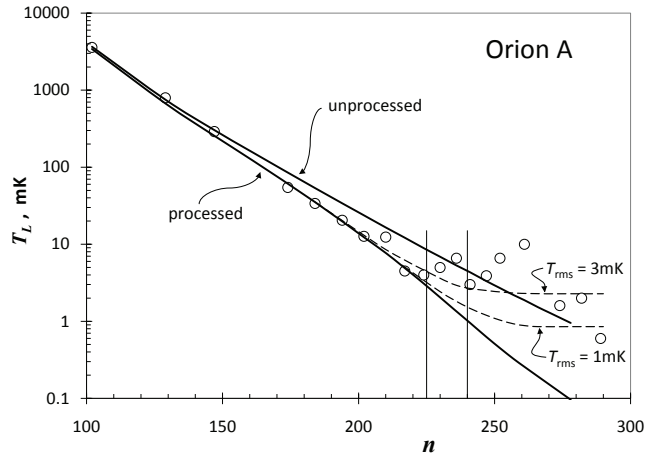}
\caption{Line temperature, $T_{L}$, vs. $n$ for the Orion Nebula at 6
GHz. Open circles represent line temperatures obtained by BASV: processed
line temperatures after six FS-overlaps (\citet{Belletal2011}). The
theoretical curves result from \citet{Lockman&Brown1975} model. Upper
and lower solid curves show unprocessed and processed line temperatures
from the model spectra without noise. Upper and lower dashed curves
show processed line temperatures from model spectra with 3 and 1~mK
noise, correspondingly. Vertical lines are positioned at $n$ values
corresponding to theoretical processed temperatures $T_{L}=3\:\textrm{mK}$
(left) and $T_{L}=1\:\textrm{mK}$ (the right vertical line). This
plot indicates that our model of processed line temperatures agrees
with BASV's findings for $(n,\Delta n)\lesssim(224,11)$. Above this
limit, our model suggests BASV's results are dominated by noise fluctuations.}

\label{fig:T_L_processed_versus_n_orion_6ghz} 
\end{figure}

\begin{figure}[h]
\includegraphics[bb=130bp 60bp 797bp 530bp,clip,scale=0.380]{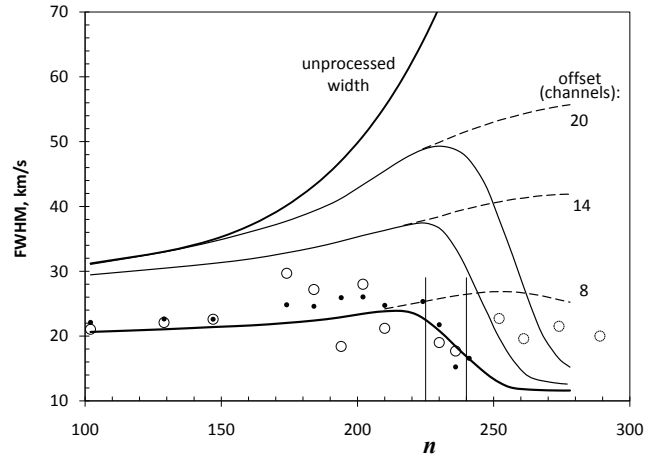}
\caption{Processed width (six FS-overlaps) vs. $n$ computed for different FS-offsets
(8, 14, and 20 channels) for the Orion Nebula model at 6 GHz. Dashed
curves---no noise; solid curves---1 mK noise added to model spectra.
Open and filled circles show the processed widths obtained by BASV
for the Orion Nebula and W51. The upper thick solid curve shows the
unprocessed width vs. $n$. Vertical lines are positioned at $n$
values corresponding to theoretical processed temperatures $T_{L}=3\:\textrm{mK}$
and $T_{L}=1\:\textrm{mK}$ (see Figure \ref{fig:T_L_processed_versus_n_orion_6ghz}).
The lower thick solid curve corresponds to BASV's eight-channel FS-offset
and 1 mK noise fluctuations. It agrees well with BASV's processed
findings for $(n,\Delta n)\lesssim(241,14)$. In the transition zone
(between the two vertical lines), the modeled width behavior imitates
\textquotedblleft{}narrowing\textquotedblright{} of the processed
spectral lines reported by BASV.}

\label{fig:width_processed_versus_n_and_offset_orion_6ghz} 
\end{figure}

Figures \ref{fig:width_processed_versus_n_and_overlaps_orion_6ghz}--\ref{fig:width_processed_versus_n_and_offset_orion_6ghz} show the model results for processed line temperatures and processed widths
versus principal quantum number, $n$, for the Orion Nebula model
at 6 GHz. To avoid ambiguity associated with the cleaning procedure,
we process each modeled spectral line separately. The processed line
width is the FWHM of the central feature measured above the zero level
(see Figure \ref{fig:lorentz_2_4_6overlaps}). 

Figure \ref{fig:width_processed_versus_n_and_overlaps_orion_6ghz}
shows the processed widths (FWHM) for 2, 4, and 6 FS-overlaps (1, 2,
and 3 cycles). Following BASV, we used six overlaps and an offset of
625 kHz = 31.2 km s$^{-1}$, which corresponds to eight spectral channels (channel
width of 78.13 kHz = 3.9 km s$^{-1}$). Noise was not added to the model spectra
when computing these curves. The dashed line shows BASV's FS-offset
(eight channels = 31.2 km s$^{-1}$). After six FS-overlaps,
the processed width (FWHM) remains at the level of $\sim\!\!70\%,\,\ldots,\,85\%$
of the offset, reaching a maximum at $n\simeq260$, then drops down
to 81\% at higher $n$. This behavior reflects a growing contribution
to the line shape from the nebula's low-density outer layer at very
high $n$. The greater contribution of the outer low-density layer
at high $n$ also explains the deviation of the reduction factors
(crosses) in Figure \ref{fig:amps_widths_6_overlaps} from the ``Lorentzian''
behavior.

Figure \ref{fig:T_L_processed_versus_n_orion_6ghz} shows the line
temperature, $T_{L}$, versus $n$ for the Orion Nebula. Open circles
are BASV's processed $T_{L}$. Solid curves show unprocessed (upper
curve) and processed $T_{L}$ for the model spectra without noise.
Dashed curves show processed $T_{L}$ for model spectra with 1 and
3~mK noise added. These curves are obtained using the following Monte
Carlo method. For each modeled spectral line feature, normally distributed
noise is added, then FS-overlapped six times. $T_{L}$ and $w$ of this
processed feature are then estimated by a Gaussian fit, as done by
BASV. The least-squares-fitting algorithm searches for the best fit
within a $\pm2$ channel interval about the central line frequency.
These steps were trialed 1000 times to determine the mean and standard
deviation values of $T_{L}$ and $w$ for each processed spectral
line feature. This method was carried out for the expected 1~mK noise
rms and, additionally, a 3~mK noise rms. Two vertical lines in Figure
\ref{fig:T_L_processed_versus_n_orion_6ghz} are positioned at $n$
values corresponding to the processed, noise-free model temperatures
$T_{L}=3\:\textrm{mK}$ (left line) and $T_{L}=1\:\textrm{mK}$ (right
line). Given BASV's $\, T_{{\rm rms}}\sim1\:\textrm{mK}$, the vertical
lines represent the $3\sigma$ and $1\sigma$ limits. The corresponding
limiting values of $n=224$ ($\Delta n=11$) and $n=241$ ($\Delta n=14$)
are also indicated by vertical lines in Figure \ref{fig:width_processed_versus_n_and_offset_orion_6ghz}. 

Processed widths versus \emph{n} for 8, 14, and 20 channel offsets
are shown in Figure \ref{fig:width_processed_versus_n_and_offset_orion_6ghz}
for the 6 GHz Orion Nebula model. For comparison, the thick upper
solid curve is unprocessed noise-free model widths. Dashed curves
in Figure \ref{fig:width_processed_versus_n_and_offset_orion_6ghz}
correspond to processed noise-free model widths, while solid curves
are the corresponding processed widths resulting from the Monte Carlo
method described above using 1 mK noise added to the model spectra.
Open and filled circles in Figure \ref{fig:width_processed_versus_n_and_offset_orion_6ghz}
are BASV's processed widths for the Orion Nebula and W51 (see also
\citet{Belletal2011}). The left and right vertical lines show the
$3\sigma$ and $1\sigma$ confidence limits from Figure \ref{fig:T_L_processed_versus_n_orion_6ghz}.

\section{Discussion}

The multiple frequency switching technique reduces amplitudes of broad
baseline variations and therefore can be used to systematically ``flatten''
baselines (\citet{Bell&Feldman1991,Belletal1992,Belletal1993}). Compared
to standard spline/polynomial baseline fitting, which allows for subjective
decisions, the FS technique reduces baseline variations objectively.
For spectral features that are much wider than the offset, $h$, FS
can be likened to differentiation, which acts as a high-pass filter
in the frequency domain (see, e.g., \citet{Hamming1998,Owen1995}).
In this case, six overlaps approximates the sixth derivative with respect
to frequency, which makes FS an efficient high-pass filter. This not
only reduces broad variations in the spectrum, but in BASV's case
(six overlaps, $h<w$ ), it also reduces broad spectral line wings larger
than the FS-offset, which manifests as line narrowing (see, e.g.,
Figure \ref{fig:lorentz_2_4_6overlaps}). 

We use the \citet{Lockman&Brown1975} model of the Orion Nebula and
simulate the conditions of BASV's observations at 6 GHz: telescope
beam size, frequency range, channel width, offset, number of overlaps,
and noise temperature. When modeling, we can avoid some difficulties
that BASV faced when they processed the real spectrum. The real spectrum
is abundant with spectral lines; after overlapping six times, the
resulting spectrum is crowded with lines and reference lines. This
situation is further complicated when significant noise fluctuations
are present, which is BASV's case when $\Delta n>11$, so cleaning
becomes a non-trivial and subjective procedure. In our case cleaning
is trivial as we can model and clean each RRL individually, with and
without noise. 

For all $\textrm{H}n\Delta n$ lines studied in BASV, the unprocessed
width (i.e., before FS) is greater than the FS-offset used. We show
that the processed width weakly depends on $n$ and $\Delta n$ (see
Figure \ref{fig:width_processed_versus_n_and_overlaps_orion_6ghz},
the six FS-overlaps case). In fact, it remains about $0.8\, h$ for
a wide range of $n$ and $\Delta n$. This property of the FS technique
makes it insensitive to changes in RRL widths when $h<w$. (It is
for this reason that we justify the inclusion of BASV's processed
widths for W51 at 6 GHz in Figure \ref{fig:width_processed_versus_n_and_offset_orion_6ghz}.)
The FS technique can be used to systematically eliminate broad baseline
variations to detect weak lines, however, in BASV's case (six overlaps,
$h<w$), this technique is not well suited to study RRL broadening.
If one's purpose is to test the theory of spectral line broadening,
the choice of the offset made in BASV ($h<w$) is not optimal. 

It is shown in the previous sections that FS reduces the S/N (see Equation
(\ref{SNRSNR})) and spectral line amplitudes (Figure \ref{fig:amps_widths_6_overlaps}).
As a result of amplitude reduction, the processed line temperature
drops rapidly to the noise temperature $T_{\textrm{rms}}=1$ mK as
$n$ and $\Delta n$ grow (the lower solid curve in Figure \ref{fig:T_L_processed_versus_n_orion_6ghz}).
The model processed line amplitude reaches 3 mK ($3\sigma$ level)
at $n=224$ ($\Delta n=11$) and 1 mK ($1\sigma$ level) at $n=241$
($\Delta n=14$). The interval between $n=224$ and 241 (between $\Delta n=11$
and 14) is a transition zone between observable spectral lines and
noise. Figure \ref{fig:width_processed_versus_n_and_offset_orion_6ghz}
shows that the processed width behavior in the transition zone\emph{
}imitates ``narrowing'' of the processed\emph{ }spectral lines observed
by BASV in spectra of the Orion Nebula and W51 at 6 GHz. This ``narrowing''
is the result of ``forcing'' a program to fit Gaussians to features
dominated by noise fluctuations---the procedure used by BASV. 

Spectra presented in \citet{Bell1997}, BASV and \citet{Belletal2011}
result from the overlapping and cleaning steps explained in Section
3. Though BASV applied overlapping to the entire spectrum, cleaning
was applied selectively to features located near the expected spectral
line frequencies, even if the processed line temperatures were below
$3\,\sigma$. This subjective and selective approach could create
``spectral lines'' from random fluctuations at the $\lesssim3\,\sigma$
level, which are abundant in the spectrum.

For $\Delta n\geq14$ the width errors predicted by our Monte Carlo
simulation are comparable to the widths themselves. Therefore, BASV's
data points for $\Delta n\geq14$ shown in Figures \ref{fig:T_L_processed_versus_n_orion_6ghz}
and \ref{fig:width_processed_versus_n_and_offset_orion_6ghz} are
likely to be misinterpreted noise features. 

In conclusion, FS is a useful technique for detecting and measuring
weak spectral features, if required corrections are minimal, that
is if the FS offset is greater than the line width: $h\gtrsim w$.
Based on our simulation results, we argue that BASV's $T_{\textrm{rms}}\approx1$
mK observations of hydrogen RRLs from the Orion Nebula at 6 GHz are
limited to $\Delta n<14$. Within this limit, we find good agreement
between their results and a Monte Carlo simulation based on conventional
Stark broadening theory and, therefore, we argue that BASV's findings
do not necessitate a revision of RRL Stark broadening theory. We suggest
that further tests of line broadening theory above this limit will
require observations with sub-mK sensitivity and improved baseline
stability. 

\acknowledgements{We thank Morley Bell for helpful communications when learning about
FS. We thank Johannes Buchner for enlightening discussions about FS
and various software tools. We thank Ron Maddalena for assistance
with NRAO's 140 ft data archive. Finally, we thank Miller Goss for
many encouraging and valuable discussions.}
\bibliographystyle{apj}
\bibliography{bibliography1}

\end{document}